# A NOVEL HARMONIC KLYSTRON CONFIGURATION FOR HIGH POWER MICROWAVE FREQUENCY CONVERSION

**Alberto Leggieri[1-2], Mostafa Behtouei[3], Graeme Burt[4],**

**Valery Dolgashev[5], Franco Di Paolo[2], Bruno Spataro[3]**

[1] Thales Microwave and Imaging Subsystems, Vélizy Villacoublay, France;

[2] Dipartimento di Ingegneria Elettronica, Università di Roma Tor Vergata, Italy;

[3] Istituto Nazionale di Fisica Nucleare - Laboratori Nazionali di Frascati, Italy;

[4] The Cockcroft Institute, Daresbury, Warrington, UK;

[5] SLAC National Accelerator Laboratory, Menlo Park, US.

## ABSTRACT

A new frequency converter, operating at significantly higher power and efficiency than previous devices, is described in this paper. The proposed device is implemented as a klystron structure where a new design principle is used. New analytical formulas and a specific design procedure are proposed. The klystron frequency multiplier can be suitable for telecommunications and non-lethal weapon, scientific and medical particle accelerators while the most interested exploitations are in the field of high gradient particle acceleration and FEL devices for which no performant sources exist. The advanced klystron multiplier can replace all the low level circuitry for frequency multiplication as a less expensive alternative. Efficiencies in the range of 60% in the K-band range with power levels of 30 MW are possible without phase noise, sideband generation, jitter or chirp effects. The presented design principle is applicable to other bands or power levels.

## 1. Introduction

As RF and mm-wave amplifiers move to higher frequencies it becomes harder to generate high driving powers and to measure, process and modify the drive signal in order to control the output. In order to simplify the control loop the signal is often down-converted to lower frequencies for manipulation before up-converting back to the operating frequency. At the same time the efficiency and gain of high power amplifiers decreases as the frequency increases, requiring larger RF drive and DC power. Electron tubes [1] [2] [3] [4] [5] and solid-state devices [6][7][8][9] have several limitations: image sidebands, phase noise, jitter and chirp. In addition to the amplification, the proposed structure converts the intermediate frequency signal to RF signal allowing circuitry to be removed. Recently klystrons have been developed with higher efficiency however klystrons that generate higher harmonics suffer from very low efficiency (max 23%).

In the frame of the Compact Light XLS Project [10], a design concept for high efficiency multiplier based on Klystron structure is proposed. The device is based on a configuration composed by both reentrant and pillbox X-band and Ka-band resonators operating in the $TM_{010}$ mode at the first, second and third harmonic of the input signal. In traditional structures, harmonic cavities produces bunches with a slow convergence in the bunch center while creating stronger convergence of electrons at the bunch edges: as part of standard klystron design methodologies BAC method [1], COM [2] or CSM [3] and in superior harmonic klystrons [4]. In the Klystron-multiplier, the harmonic cavities closer to the output structure produces convergent bunches in order to maximise the harmonic content at multiplied frequency. Unlike traditional Klystrons [11], in the proposed structure, the velocity modulation is arranged to create a higher harmonic current modulation [12]. A new design approach is proposed in this paper. The design considers the set of frequencies of a coupled cell oscillator used as output coupler. Differently from an extended interaction klystron [5], one frequency is used for the output signal and a second one is used to superpose a second normal mode to create a particular condition that limits electron reflections and increases efficiency. The latter frequency is generated by a dedicated harmonic gain resonator installed before the output coupler. In order to obtain the harmonic gain, an opportune design strategy and closed form formulas are proposed.

An extreme frontier application is chosen as case study to illustrate the design process capabilities and numerically verify the design technique. Without need of redesign or adjustment, but only by adjusting the power input, the proposed case study structure can operate below or above the thresholds of traditional parameters for RF vacuum breakdown used as design criteria for normal conducting high gradient structures. This strategy allows to use the same design for investigate different technologies including materials and manufacturing techniques, giving the possibility to investigate the limit of new technologies.

## 2. Principle of operation of the proposed device

The proposed theory starts from the traditional kinematic analysis in a two-cavity klystron to estimate the bunching process in case of a low depth sinusoidal modulation of charge velocity as follows. By considering an electric space charge leaving the input cavity at the instant $t_1$, the modulated current arrives in the output cavity, after have traversed the segment $l$, in the instant $t_2$ defined, in the phase space, as $\omega t_2 = \omega t_1 + \theta_0 - X \sin(\omega t_1)$. The quantity $X$ the bunching parameter that individuates the effect of charge grouping while the $M$ is the beam coupling factor, that individuated the effective portion of the oscillating gap field experienced by the moving electron defined as $X = M V_{RF} \theta_0/(2V_0)$,

$$X = M \frac{V_{RF}}{2 V_0} \vartheta_0 \qquad (1)$$

where $V_0$ is the beam voltage, as the beam leave the electron gun, and $V_{RF}$ is the shunt voltage at the input cavity oscillating at the electrodynamic field pulsation $\omega$. The value $\theta_0 = \omega l / u_0$ is the transit angle that represents the traveled distance in the phase space, where $u_0$ is the velocity at which the electron beam enters the first cavity. The quantity $M$ is the beam coupling factor, that individuated the effective portion of the oscillating gap field experienced by the moving electron [11]. It can be numerically calculated by integrating the electric field resulting by an Eigen mode simulation by discretizing the field along z-steps in the following formula:

$$M = \frac{\left|\int_0^L E_z(z) e^{j\frac{2\pi}{\lambda}\frac{z}{\beta(z)}} dz\right|}{\left|\int_0^L E_z(z) dz\right|} \qquad (2)$$

Where $L$ is the gap length where to integrate, $E_z(z)$ the value of the axial electric field in function of the axial coordinate z, λ is the wavelength for the considered cavity mode (in general $TM_{010}$) and β is the velocity factor $u_0/c$, where $u_0$ is the velocity of electron (even in presence of a velocity dispersion, this value can be assumed as the velocity at the beam pipe inlet).

The $n^{th}$ harmonic component of the current is $I_n=2I_0J_n(nX)$ where the $J_n(nX)$ is the Bessel functions of the first kind for the order $n$ calculated in the point $nX$, in which $I_0$ the DC beam current. Since the velocity across the structure is dispersed around this value due to the velocity modulation, for a negligible transit angle, the maximum power provided to the load by the electron beam is $P_n=I_0 V_0 J_n(nX)$, considering the potential reduction. Hence the conversion efficiency is $\eta_n= J_n(nX)$. It can be noted that, the bunching parameter $X$ decrease when the harmonic order $n$ increase following a linearized law $X \approx 1+0.81n^{-2/3}$ and the efficiency slowly decrease as $\eta_n= J_n(nX) \approx 0.65n^{-1/3}$[12]. The modulated current arriving in the second gap in the instant $t_2$, after have traversed the segment $l$, can be expressed as the sum of all harmonic contributions produced in all the time segments $t_{11}$, $t_{12}$,… $t_{1n}$ as $I/I_0 = \sum_{n=1}^n 1/[1-X\cos(\omega t_{1n})]$ [11]. The bunch can be expressed as the space charge modulated current density in function of $t_1$. By considering a small amplitude modulation of the shunt voltage $V_{RF}$ over $V_0$, where when $V_{RF} << V_0$, the bunch parameter assumes the value $X = 1$, in that case the bunch has an exponential shape meaning that it became to be rich in harmonics. On contrary, $X = 0$ when $V_{RF} = V_0$ meaning no bunching occur and the current remains constant. A pure linear modulation occur for $0 \le X < 0.5$ where the distribution is sinusoidal and higher order harmonics are negligible; hover the same happens to the power conversion efficiency

The $n^{th}$ harmonic component of the modulated current can be expressed as an integral function of the oscillations at $n\omega t_2$ [12] as well as a bilinear function of the single oscillation at $\omega t_1$ [11]. By combining the two relations, the harmonic current depth that represents the portion of the DC current transferred to the $n^{th}$ harmonic, can be defined as $|I_n((\omega t_1, \omega t_2))|/I_0 = |_0\int^{2\pi} e^{jn\omega t_2} d\omega t_1 = 1/|1-X\cos(\omega t_1)|$. The local maxima and minima of $X$ closer to the inflection discontinuity where $\omega t_1 = \omega t_2$, identify two stationary phase points that occurs when the first derivative $d\omega t_2/d\omega t_1 = 0$ and are comprised in the phase space interval where $0 < \omega t_1 < \pi/2$ that can be defined as $\omega t_0 = \omega t_1|_0^{\pi/2}$. If the transit angle $\theta_0 > d^3\omega t_2/d^3\omega t_1$, the condition $d\omega t_2/d\omega t_1 = 0$ is satisfied for $\omega t_1 = \omega t_-$ and $\omega t_+$ that provide the phase space width $\omega t_0 = \omega t_+ - \omega t_-$, where the stationary phase points are confined. By defining $\omega t_{++} = d^2\omega t_2/d^2\omega t_1|_{\omega t_1 = \omega t_+}$ and $\omega t_{--} = d^2\omega t_2/d^2\omega t_1|_{\omega t_1 = \omega t_-}$, the harmonic current depth, expressed in function of $\omega t_2$, $X$ or $\theta_0$ as $|I_n(\theta_0)|/I_0$, oscillates around $<|I_n(\theta_0)|/I_0> = \min\{|I_n|/I_0\}/2 + \text{Max}\{|I_n|/I_0\}/2$, as

$$\begin{cases} Max\left\{\frac{|I_n\theta_0|}{I_0}\right\} = \sqrt{\frac{2}{\pi n}}\left[j\sqrt{\frac{1}{\omega t_{++}}} + \sqrt{\frac{1}{\omega t_{--}}}\right] \\ Min\left\{\frac{|I_n\theta_0|}{I_0}\right\} = \sqrt{\frac{2}{\pi n}}\left[j\sqrt{\frac{1}{\omega t_{++}}} - \sqrt{\frac{1}{\omega t_{--}}}\right] \end{cases} \quad (3)$$

In case of small modulation depth, where $V_{RF} << V_0$, the quantities behaves as $\omega t_- \approx -\omega t_+$ and $\omega t_{--} \approx -\omega t_{++}$ and, because of this symmetry, the oscillations of $|I_n|/I_0$ are maximized. When the shunt voltage at the input cavity becomes $V_{RF} = V_0$, a portion of the modulated current is reflected from the output structure and only the maxima persist for any value of $X$; hence the peak-to-peak amplitude of $|I_n|/I_0$ is halved. This effect results in a strong limitation of conversion efficiency due to reduction of the bunching parameter while modulation depth increases.

In order to overcome this limit, the following solution is proposed: In the bunching circuit, the full oscillation of the harmonic current can be preserved if the multi cavity klystron, composed by $N_{cav}$ resonators showing RF shunt voltages $V_{RF} = V_{RF\ 1}, ..., V_{RF\ N_{cav}}$ exhibit $V_{RF\ i} << V_0$ until the last gain cavity ($N_{cav}$-1). This condition can be satisfied if the shunt voltages grow slowly along the structure and grow dramatically in the output sector, for example and exponential increase along the structure. However, in the output structure, we do not limit RF field intensity as this would limit the output power and it shall be designed in to maintain the full oscillation of the harmonic current. Preventing these reflections means to maintain the phase space symmetry. The full oscillation of $|I_n|/I_0$ can be preserved if, before the output cavity, the structure exhibits a sector where the modulated beam current is forced to remain constant and electrons are prevented from being reflected. This can be achieved by splitting a multi-gap output cavity into two sections where the first sector serves to implement the constant beam current and the second sector to extract the energy from the beam. In the first section the beam current is forced to remains constant and electrons are prevented to be reflected by the large modulation depth $V_{RF}>>V_0$; at the same time the kinetic energy, stored in the RF field by the interaction of the second section, is prevented to be reinjected into the beam  By considering a charge having traversed the first sector, in the second sector the maximum beam kinetic energy can be extracted without re-acceleration if the flight time of the bunch in the interaction space corresponds at least to a half period at the  desired output harmonic. . In the first sector the superposed modes shall oscillate in phase opposition in a narrow band close to the output frequency at which the modulated beam current is forced to remain constant and electrons are prevented from being reflected. Pill-box cavity cells are suitable for the output structure. With the approximation that the velocity is dispersed around the initial value, in order to allow a  charge to traverse the output structure in a period of the harmonic frequency, the output structure should have a total interaction length $L'_{out}$ that is longer than the beam wavelength at the desired multiplied frequency $L'_{out}=\lambda_e=u_0/f_{out}$. This condition can be achieved if the gap-to-gap distance $\Delta z$ and the gap width $d$ of the coupled cells of the output structure are equal, thus $\Delta z'=d'_{\text{output structure}}=L'_{out}/N_{cell}= u_0/(f_{out}/N_{cell})$. If high beam voltages are used where $u_0/c>½$ (this is the convenient case for a high efficiency multiplier) the interaction region length can be a quarter wavelength of the RF output frequency. Since it is not possible to create a gap longer that the gap-to-gap spacing inside a multi-cell oscillator, the limit condition imposes that the gap spacing is at least equal to the gap width: $\Delta z = d_{\text{output structure}} = L_{out}/N_{cell}= c/(f_{out}/N_{cell})$; then they can be numerically adjusted. If the constant current sector is composed by $N_{const}$ cells, the interaction sector is composed $N_{int}$ cavities. An equal number of cavities to the two sectors is assigned.

If the beam voltage is strong enough to allow the particle to cross a gap of length equal to a quarter free-space wavelength of third harmonic frequency while not exceeding a half period ($\lambda/4 < /2 \cdot u_0/c$), mixed accelerations and decelerations can be prevented in the gap and the interaction can be maximized. In the case of $u_0/c<½$, the interaction efficiency will be sensibly lower than in the case of $u_0/c>½$, hence high voltage guns are suggested. By defining $N_{harm}$ as the harmonic order of the desired output signal, the gap $d$ of the cells output cells is

$$d = \begin{cases} \frac{u_0/c}{N_{harm}}\frac{\lambda}{N_{Cell}} & \text{if } \frac{u_0}{c} \leq ½ \\ \frac{1}{N_{harm}}\frac{\lambda}{N_{Cell}} & \text{if } 1 > \frac{u_0}{c} > ½ \end{cases} \quad (4)$$

This behavior can be obtained by a system of four cells oscillating in the $2\pi/3$ at the desired output frequency $f_{out}$ where the two sectors each have 2 cells. In this case, the gap-to-gap distance and the gap width shall match the quarter wavelength of max resonance frequency $\Delta z=d_{\text{output structure}}=\lambda_{out}/4$, and the cells are dimensioned as a beam quarter wavelength. The output structure operates in dual mode, by superposing the $2\pi/3$ mode and the $\pi$ mode: The frequency of the $2\pi/3$ mode determines the output frequency of the total structure ($F_{2\pi/3} = F_{out}$) and that of the π mode determines the space charge harmonic component that is submitted to the high gain ($F_\pi$ = $F_{\text{space charge gain}}$). A suitable solution where the output structure is composed by a system of four cells implies:

$$d = \begin{cases} \frac{u_0/c}{N_{harm}} \frac{\lambda}{4} & if \frac{u_0}{c} \leq \frac{1}{2} \\ \frac{1}{N_{harm}} \frac{\lambda}{4} & if\ 1 > \frac{u_0}{c} > \frac{1}{2} \end{cases} \quad (5)$$

Once defined the output structure, the π mode frequency is used to define the oscillation frequency of the last gain cavity. The output cells should have the same the R/Q and M as the bunching circuit.

In this output coupler, four Eigen-modes can be superposed and exploited: The constant behavior of the current, in the first half of the output structure, is limited by the frequency separation of the two modes *π* and *2π/3*, which produces a residual modulation of the current by an additional compression and decompression pattern along the bunch frequency spectrum. These two modes in the output structure should have a beam-coupling factor close to that of the uncoupled cavity and ideally a "good" coupling (*M≥0.6*) while the other normal modes should not couple to the beam (*M≤ 0.2*).

$$\begin{cases} M_\pi \approx M_{2\pi/3} \geq 0.6 \\ M_0 \approx M_{\pi/3} \leq 0.2 \end{cases} \quad (6)$$

The resonance frequencies of the output structure are calculated by the Eigen-mode solution of the coupled oscillator with the condition to have the same field components at the two extremities to consider the effect of the drift tubes [13].

$$\begin{cases} \omega_q = \frac{\omega_0}{\sqrt{1+k \cos\left[\frac{q\pi}{N_{Cell}-1}\right]}} \\ \alpha = q \frac{\pi}{N_{Cell}-1}\ ,\ q = 0,\ 1,\ 2,\ \ldots,\ N_{Cell}-1. \end{cases} \quad (7)$$

Where α is the mode order, i.e. the phase shift between adjacent cells and $N_{cell}$ is the number of cells that compose the output coupler and *k* is the cell-to-cell coupling factor that individuates the field ratio and the frequency detuning.

$$k = 2 \frac{Max\{\omega_q\} - min\{\omega_q\}}{Max\{\omega_q\} + min\{\omega_q\}} \quad (8)$$

The frequency manipulation follows the beam stratification: the higher distance from the image charge on the drift tube allows the inner electrons to vibrate more rapidly than the outermost with higher plasma frequency. The contribution to the superior harmonic content is mostly given by the inner part of the beam while the outer behaves like the fundamental. Narrow beam pipes and pillbox cavities are indicated to implement the output coupler and the harmonic gain cavities for this purpose.

The bunching circuit is composed of a fundamental (input) circuit resonating at the input frequency $f_{in}$ followed by a (output) harmonic circuit oscillating at the frequency $f_{out}=N_{harm} f_{in}$. The fundamental frequency section is similar to that of a traditional klystron with an input cavity followed by a compression stage composed of intermediate gain cavities to provide a sufficient amount of harmonic content to be used by the last gain cavity and the output structure. The harmonic frequency section is composed by: I) A non-linear bunching circuit, composed by one or more harmonic gain cavities operating in large compression and II) A multi-gap output cavity (output structure) that is implemented by a system of coupled resonators.

The requirement on the linearity and compression levels can be satisfied with an exponential growth of the gap shunt voltages along the structure, while respecting the conditions on the bunching parameter, *X*. The input cavity and low gain fundamental cavities, known as the linear circuit, operate with *X*<0.5 and *X*≤1 respectively. The low compression stage (fundamental moderate gain cavities) operate with 1<*X*<1.84 and are composed of the same number of cavities as the linear circuit. The non-linear bunching circuit begins with the drift space after the fundamental circuit, where the bunching parameter is *X*>1.84. The non-linear circuit could be composed of a single harmonic cavity.

The harmonic bunching circuit should oscillate at the frequency of a neighboring mode of the coupled cell system in the output structure. We shall show later that, the π mode frequency represents a good solution.

The last gain cavity sharply modulates the beam velocity such that in the drift space before the output structure, the beam is compressed to enrich the higher harmonic content at one of the output structure auto-frequencies. In the formulism presented here the harmonic currents are grouped into packages, where each current within the package have similar amplitudes. In the Ka-band case study, the last cavity produces a package of 1st, 2nd and 3rd harmonics at the same current amplitude.

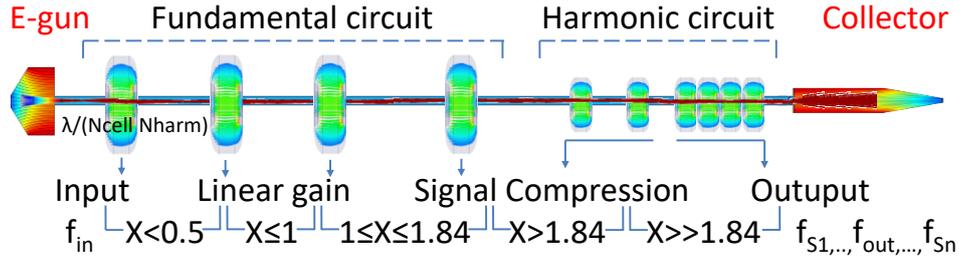

**Figure 1. Proposed frequency conversion structure.**

A prerequisite to perform this current manipulation is to allow the cavities of the harmonic bunching circuit to operate as gain cavities and not as standard harmonic cavities: In traditional structures, the harmonic cavities produce a de-bunching of the beam (divergent phases): In the proposed structure, harmonic cavities produces convergent phase beam resulting in a continuous beam bunching. No sub-bunches are created (as often happens in a traditional klystron with a harmonic cavity).

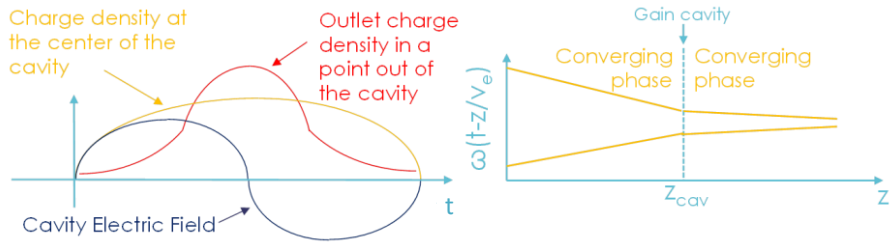

**Figure 2. Behavior of a gain cavity: Bunching (left). Applegate diagram (right).**

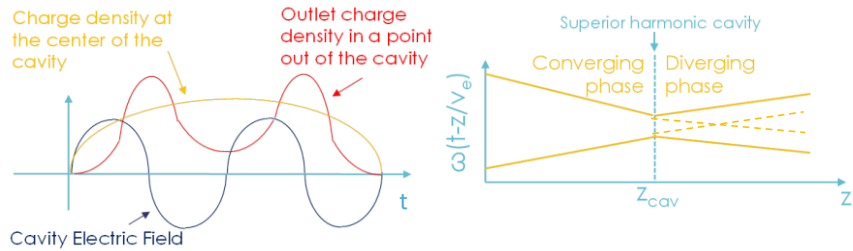

**Figure 3. Behavior of a II harmonic cavity: Bunching (left). Applegate (right).**

## 3. Design technique to synthesize the structure

In order to design the multiplier, a traditional Klystron small signal design [11] is performed with an analytical formulation that provides the closed form formulas for key design parameters. A hybrid design procedure is proposed since an analytical solution does not exist and the distance between cavities. *l,* and the *R/Q* can be calculated numerically due to the dependence on beam current $I_0$ and large signal effects [15]. The design procedure follows two steps: I) Semi-Analytic Predefinition: This is the first step is a preliminary definition of the structure following a set of proposed closed for formulas to define gap lengths and shunt voltages while attempting to respect the condition on the bunching parameters. II) Modulation Depth Correction: To adjust the Klystron parameters to obtain a certain configuration of harmonic currents while mandatorily respecting the condition on bunching parameters.

An extreme frontier application has been chosen to illustrate the design process capabilities and numerically verify the design technique. The technical requirements of the Compact Light XLS project, are used with a frequency of 36 GHz, the 3rd harmonic of the drive frequency [10]. The design of the Klystron multiplier has been performed with the aid of the code KlyC [15].

### 1.1. Semi-Analytic Definition

The operating frequencies and power levels, as well as the electron beam features, are considered as an input for the design. A high-energy beam, is preferable if possible. By following the theory reported in [16] and the procedure described in [17], an electron gun has been designed with a cathode potential $V_0$ = -480 kV to produce a beam current of $I_0$ = 100 A by relying on a magnetic focusing field of *B* = 3.2 T. As it will be discussed later, the drift tube radius has been set to $b$=1.2 mm and the beam radius is dimensioned by considering that it should be the 20% less than the drift tube; hence it is chosen to be $a$=1 mm [17]. The beam spot contains a current density $J_0 = 3.18 \cdot 10^7$ A·m$^{-2}$ that, with an initial energy $V_0$ = 480 keV, is injected in the interaction structure [16]. For the case study, in the input cavity, the inlet electron wavelength $\lambda_e = u_0/f$ = 21.4 mm, specifies a beam propagation factor $\beta_e = 2\pi / \lambda_e$ = 293.5 m$^{-1}$ while the RF wavelength is $\lambda = c_0/f$ = 25.0 mm, and the wave propagation factor $k = 2\pi / \lambda = \omega/c_0$ = 251.5 m$^{-1}$. By assuming a 1-dimensional approach (beam propagating only on-axis), the beam to wave coupling coefficient M can be assumed as the cell transit-time factor [13].

For the case study, $N_{harm}$=3 to operate at the 3$^{rd}$ harmonic of 12 GHz; since the free space wavelength at 12 GHz is $\lambda=c/f$=25mm while $u_0/c$=0.86, the (2) gives $d$=2.08 mm for the output structure and R/Q=100 Ω. The voltage configuration is defined starting from the maximum admissible electric field and the previously defined gap length with an ideal coupling factor. For the case study, a value $E_{max}$ = 300 MV/m has been chosen basing on the average value reached during recent experiments on high gradient accelerators, where electric fields between 200 and 500 MV/m have been obtained without instabilities (until the continuous wave if cytostatic structures are used) [16]. Even if this choice lead to overcome traditional design criteria for RF vacuum breakdown in high gradient normal conducting structures [18] [19], the designed structure will be aimed to operate, at slightly different power level below the traditional thresholds, giving the possibility to investigate the limit of new technologies. The case study multiplier works in 200 ns pulses with 100 Hz repetition rate. This condition gives a duty cycle $\delta$=0.05% that allows for a weighted electric field of $\delta \cdot E_{max}$ = 15.0 kV/m.

Without considering that the output structure is implemented as a multiple oscillator system, the N$_{cav}$ cavities of the klystron are numbered as follows: i=1 for the input cavity; from i=2 to i=N$_{cav}$ -1 for the gain cavities; i=N$_{cav}$ for the output structure. In order to reach the desired E$_{max}$ in the output structure, while fitting the condition on the bunching parameters, an exponential configuration of the shunt voltages V$_{RFi}$ is proposed for the i$^{th}$ cavity:

$$V_{RF\,i} = E_{Max}\,d_i\,e^{i-N_{cav}} \qquad (9)$$

The choice of cavity parameters follows the condition on the bunching parameter, $X$, should this condition not met in the predefined structure, it can be adjusted in the second step.

An equivalent bunching parameter for multiple drift spaces is proposed by assuming that the effect of the bunching between previous cavities is sufficiently low to be linearly added to the subsequent without accounting for all permutations between drift spaces. By considering that the beam energy is dispersed around $eV_0$ and the beam velocity around $u_0$, the equivalent bunching parameter [11] between the cavity $i$ and $i$-1, is

$$X_{i-1,i} = \sum_{h=2}^{i} M\ 2\pi f_{h-1} \frac{l_{h-1,h}}{u_0}\ \frac{V_{RF\,h-1}}{2\,V_0}, \qquad (10)$$

The low frequency circuit, composed by first harmonic cavities, provides a modulated beam that is moderately compressed going into the drift space after the last gain cavity by smoothly reinforcing the 3$^{rd}$ harmonic content at the output frequency $N_{Harm} f_{in}$. The external coupling of the input cavity $Q_{ext}$ is calculated numerically to reach the desired gap shunt voltage for the given input power. The distance between the input cavity and the first gain cavity $l_{1-2}$, as well as the distances between all other cavities, $l_{i-1,i}$, should be calculated numerically to respect the requirements on the shunt voltage and bunching parameter even with a maximized coupling factor.

Since the case study Klystron has been implemented with $N_{cav}$ =6 cavities (output structure assumed as single oscillator), the (3) gives $V_{RF\,1}$ = 4.2 kV and $V_{RF\,2}$ = 11.1 kV for the input and for the first gain cavity respectively. On the KlyC large signal code [15], a desired Pin=500W has been imposed and, in order to obtain the desired $V_{RF\,1}$, the input cavity has been defined with $R/Q$=110 and $Q_{ext}$ = 42. The first gain cavity has been placed at $l_{1-2}$=37 mm away from the input cavity to get the desired $V_{RF\,2}$ = 4.2 kV. At this distance, $X_{1-2}$= 0.048. To improve the bunching process, a BAC method [1] harmonic cavity has been added after the 3$^{rd}$ linear gain cavity.

As, defined before, the output structure is a system of 4 on-axis electrically coupled oscillators: For the case study, $V_{RF\,5}$ = 229.8 kV and the distance between the last gain cavity and the first cavity of the output structure has been set to $l_{5-6}$ = 17 mm, so that $X_{5-6}$ = 6.18 . The parameters of the single cavities composing the output structure are $M \approx 1$ and $R/Q$=110,the gap and gap-to-gap spacing are $\Delta z = d = 2.08$ mm and the resonance frequency is $f_{2\pi/3}$ = 36GHz. This value is analytically obtained by the Eigen-mode solution of a coupled oscillator with $N_{cell}$ =4 resonators oscillating at $f_{Uncoupled}$ = 35.50 GHz with $k$ = 0.055 considering the condition to have the same field components at the two extremities to consider the effect of the drift tubes [13]. This oscillator can be obtained with a cascade of pillbox cavities electrically coupled by central aperture of radius $b$ =1.2mm. This aperture determines the radius of the drift tube $b$. The $\pi$ mode oscillates at $f_\pi$ = 36.52 GHz and, since this is the higher frequency in the structure, the drift tube allows for a safety margin of ~2, because its TE$_{11}$ cut off frequency is $f_s$ = 73.30 GHz.

## 1.2. Modulation Depth Correction

After the predefinition, the structure is adjusted to make the first harmonic have the same modulation depth of the desired output harmonic . Note that the output cavities do not respond to the first harmonic current and, at fundamental frequency, low shunt voltages are produced). At this stage, the desired ratios between harmonic depth modulations are targeted while mandatorily respecting the requirements on the maximum desired shunt voltages, imposed by (9), and those on the bunching parameters $X$, imposed by (10). The previously defined configuration of the gap shunt voltages will be moderately modified. With the exception of the resonance frequency of the output structure and the harmonic gain cavity (in our case, $f_\pi$) the frequency detuning and the spacing between gain cavities $l$ are adjusted to obtain the required ratios between the modulation depths of the harmonic currents: $I_1 = I_{N_{harm}}$.

The velocity modulation is transformed into a current density modulation along the drift space between cavities up to the output structure. All permutations of drift spaces contribute to generate the output structure current, where electromagnetic fields resonate in double mode configuration. There is also a possibility to connect more than one waveguide in order to provide additional frequency outputs with reduced amplitudes.

In the predefined analytic structure for the case study, a balanced efficiency (the efficiency taken from subtracting the beams exit energy, including space charge power flow, and the RF ohmic losses, from the beams input energy) η=49.7% was reached, but the maximum electric field, 561 kV, was the 10% below the theoretical desired value, 624 kV and the configuration of the harmonic current modulations did not perfectly fit the desired distribution defined in (8), as the 3rd and 6th harmonic amplitudes were too high.

The desired maximum electric field has almost been obtained and the expected configuration of harmonic currents have been obtained, resulting in a higher efficiency η=54%. The comparison of bunching parameters $X$ and the modulation current depth in the harmonic circuit for the two structures is reported. The required bunching parameters and the desired modulation depth for the output harmonic have been met by a constrained modification of the shunt voltages. Two harmonic packages are created with the same amplitude: the 1st and the 3rd harmonics as well the 2nd and the 4th harmonics.

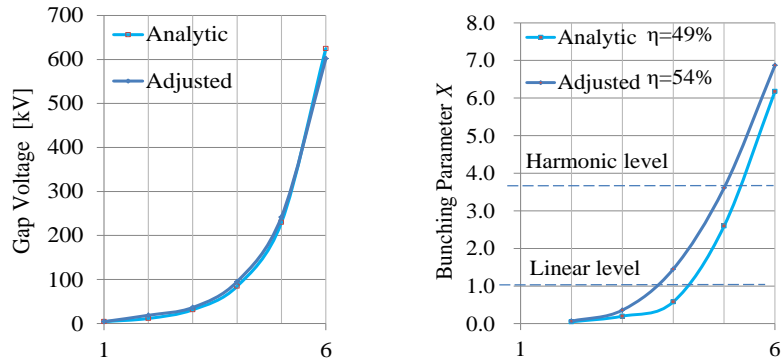

**Figure 4: Gap voltages (left) and Bunching parameters (right) of the cavities for the two structures.**

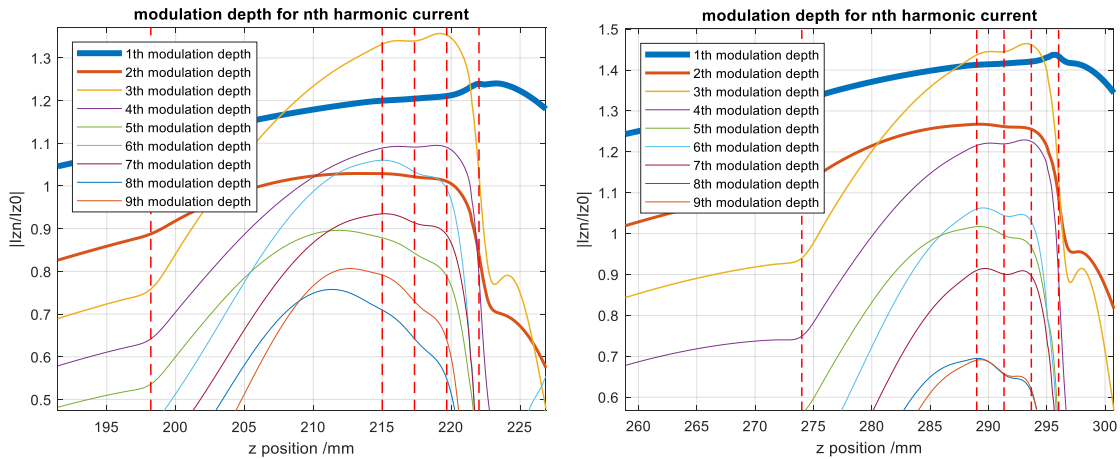

**Figure 5: Modulation depth for harmonic currents: Predefined structure (left) and adjusted structure (right). Note that the extracted power depends to the resonance frequency of the output structure that is will generate a shunt voltage at the given frequency.**

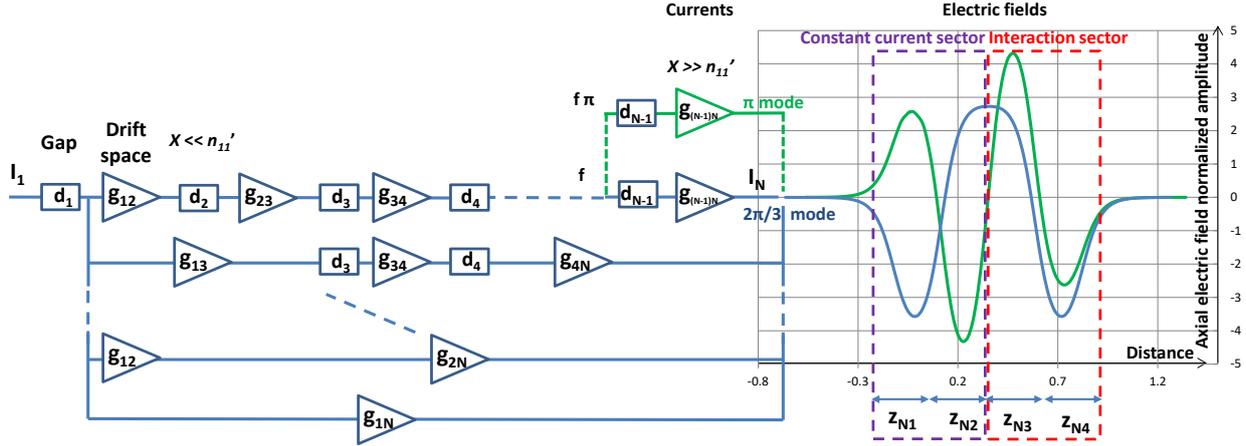

**Figure 6:** Feed-forward representation of the structure that generates the output signal at $F_{out}$ by integrating a single harmonic gain cavity (resonating at $F_\pi = F_{\text{space charge gain}}$) and a four-coupled-cells output structure (that exhibits two strong-interacting modes $2\pi/3$ and $\pi$, resonating at $F_\pi = F_{\text{space charge gain}}$ and $F_{2\pi/3} = F_{out}$ respectively).

### 1.3. Efficiency maximization

In order to maximize the interaction efficiency, harmonic cavities replace one or more cavities of the fundamental circuit. The higher harmonic cavity oscillates at an intermediate frequency whose harmonic index is $1 < N < N_{Harm}$. The modulation introduced in the new structure produces an additional modulation in the current that produces a series of harmonic voltages split into packages for the harmonic modulation depths in the output cavity. It has been observed that, for a maximum efficiency without electron reflections, the ideal configuration of modulation depths for harmonic current in the output structure has a time expression behaving like.

$$I(t) = I \sum_{n=1}^{N_{harm}} \sin(n\omega t + \varphi_n) + I \sum_{p=1}^{\infty} \frac{\pi}{N_{harm}+p} \sum_{n=pN_{harm}+1}^{N_{harm}(p+1)} \sin(n\omega t + \varphi_n) \qquad (11)$$

The position of the additional harmonic cavities and the harmonic circuit, is used to adjust the amplitudes of the harmonic currents in the output structure to match the required values as much as possible.

In the case study, the last cavity of the fundamental circuit (4th cavity, operating at 12484 MHz as shown in Fig.9) is replaced by a harmonic circuit. In the study with, $N_{Harm}=3$, the first package of harmonic currents is composed by the first three harmonic currents ($N_{harm}=3$) which have almost the same amplitude *I*, while the current phasor of the next package, 4th, 5th, and 6th harmonics, would be $I·\pi/4$ and those of the 7th, 8th, and 9th harmonic would be $I·\pi/5$. By altering the resonant frequency of the output circuit one of the 1st, 2nd or the 3rd harmonics can be chosen to be extracted with a power amplitude *P,* or for extracting a harmonic in the 2nd package a slightly reduced amplitude $P·\pi/4$ is extracted while the 3rd with an amplitude of $P·\pi/5$. Hence, the amplitude of the second and third packages should be the 78% and 63% of the first respectively. For the case study; the harmonic resonator has been implemented by a 2nd harmonic low shunt impedance pill-box cavity. The resultant structure is composed of 10 cavities. A tradeoff between reducing the beam voltage while maintaining a high power output and hence efficiency has been performed, while keeping the minimal electron velocity below -0.1c in order to avoid reflected electrons. It is possible to use a low $Q_{ext}$ alongside a strong input signal c to reduce any energy spread between electrons located at different radial coordinates of the beam. In the case study, the parameters of the input cavity have been adjusted to $d$ = 2.2 mm, $M$ = 0.962, $Q_{ext}$ = 48 and $R/Q$ = 109.5 Ω and with $V_{RF\,2}$ = 4.51 kV. At *l*=63.5 mm from the input cavity, the bunching parameter of the beam is $M$ = 0.961 and hence the center of the first gain cavity is positioned there.

The contribution to the higher harmonic content is mostly given by the inner part of the beam due to the effect referred to as beam stratification caused by both the radial variation in the space charge forces and the radial variation of the gap voltages. The choice on the cavity profiles constructively contributes to the frequency multiplication process. Reentrant cavities are indicated for input cavity and gain oscillators: They operate more intensively on the outer radial part of the beam, which is poor in higher harmonics, and it needs to be more compressed in the drift space. At the same time, this circuit segment weakly operates on the inner part of the beam; which is already rich in higher harmonic content.

For the case study, all the gain cavities, including the input cavity (1st, 2nd and 3rd) are reentrant cavities. The 4th cavity is an ordinary 2nd harmonic cavity tuned at 23.89GHz used to increase the beam density (operating on the head and tails of the bunches), while the 5th cavity is detuned at 12.28 GHz; both are low shunt impedance pill-box resonators. In these cavities, the effect to mostly operate on the outer beam section is enhanced to predispose the outer cross section of the beam to the acquisition of the higher harmonic content. The 6th cavity is the most important element of the Klystron Multiplier: it is a pill-box high shunt impedance cavity operating at the 3rd harmonic of drive frequency. The output structure, placed after the 6th cavity, consists of a system of four 3rd harmonic cavities with gap lengths and gap-to-gap spacing optimized to $\Delta z = d = 2.34$ mm and $M = 0.693$ electrically coupled with $k = 0.055$, analytically calculated as described in (8) [13]. The analytical Eigen-frequency of the π mode has been attributed to the 6th and final gain cavity to give gain at the output frequency (chosen to be the π mode frequency of the output resonator). The first cavity of the output structure has been placed at $l = 18$mm from the last gain cavity, where a nonlinear current with bunching parameter $X = 6.60$ induces a maximum voltage $V_{RF\ max} = 655.9$ kV at the output cell. The total structure is 2947.02mm long.

After the semi-analytical design and numerical optimization, a final high efficiency 10-cavity structure has been obtained. The first three cavities operate a linear bunching, while the 4th cavity reduces the convergence of the beams in phase (like in the BAC technique [1]), rather than increase the gain. Before extracting the harmonic content from the ideally prepared beam, the 3rd harmonic phasor has been amplified by the last gain cavity which operates equally over all of the beams cross section to produce the final current at the desired output frequency with both the 12 and 36 GHz currents with the same amplitude. The harmonic ratio of the space charge current is obtained with the average value of the currents ion the four output cavities normalized to the amplitude of the 1st harmonic. A more precise estimation can be expressed for each cavity. The waveform can be reconstructed by anti-transforming the sum of components. Due to the distribution in packages of the modulated beam current, several possibilities for power output are offered. Finally, the exact Eigen-frequencies of the output coupler obtained by the Eigen-mode simulation (shown in Table 1) are used, instead of the analytically calculated frequencies. The radius of the output structure has been adjusted with the software Poisson Superfrish [11] to operate in the 2π/3 mode exactly at $f_{2\pi/3} = 36.00$ GHz. With the given profile, it exhibits a π mode frequency $f_\pi = 36.64$ GHz; which is reasonably close to the theoretical value found with the analytical formula (6) by fixing an unperturbed frequency of 35.50 GHz and the same $k$.

## 4. Simulations Results

The electron beam is dramatically decelerated in the last cavity of the output section; the electrons with the largest deceleration temporarily obtain a negative velocity but are then reaccelerated in a bouncing process. Several different current modulation schemes are possible to implement a different frequency output. The case study structure, dedicated to the Compact Light XLS project, presents an efficiency of 62% and a gain of 47.8 dB with about 5% of ohmmic losses. The Output power is 29.85 MW at the 3rd harmonic of drive frequency. The max gap voltage is 647.1 kV, which is produced in the penultimate output cavity (9th cavity) which also results in the max electric field (295.6 MV/m) due to extreme bunching before energy extraction. The minimum velocity level is -0.082·c, which respects the requirement on $v_{min} > -0.1c$. No monotron oscillation and reflected electron have been detected. A numerical optimization of the frequencies of the harmonic structure can be also performed to consider both beam loading and the external quality factor of the waveguide connected to the output coupler. The presented simulations have been obtained by setting the following KlyC parameters [15]: Number of beams: 8, Space charge field order: 12, Electron wavelength division number: 256, RF wavelength division number: 128, Iteration residual limit:$10^{-4}$, Max iterations: 64, Iteration relaxation: 0.25. Due to the intrinsic low perveance, the proposed design can be implemented as a multi-beam klystron [20] using up to 10 beams at 100A, to provide ~300 MW output at 12 GHz, 24 GHz or 36 GHz, while maintaining the 60% range of efficiency, as well as 152 MW at 108 GHz by adopting a 9th harmonic circuit.

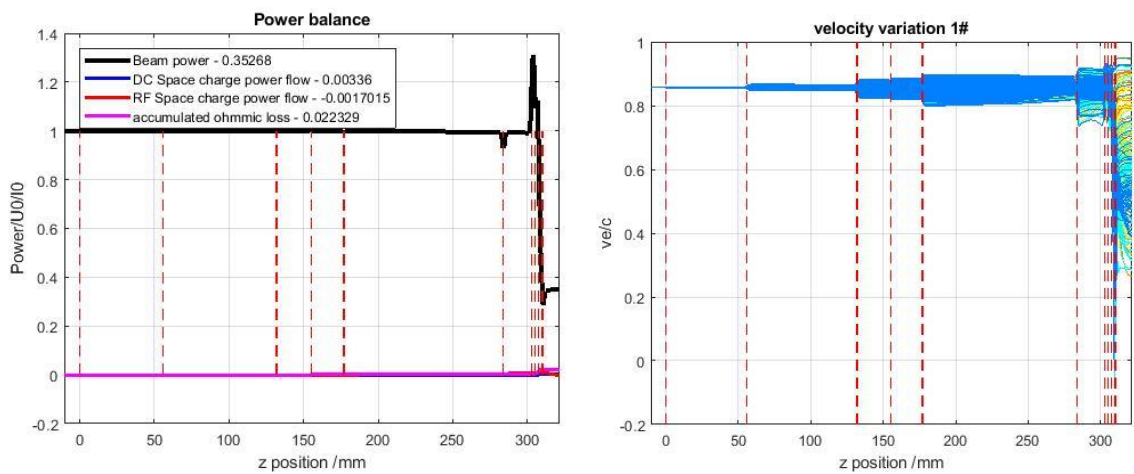

Figure 7: Power balance (left) and Velocity variation (right).

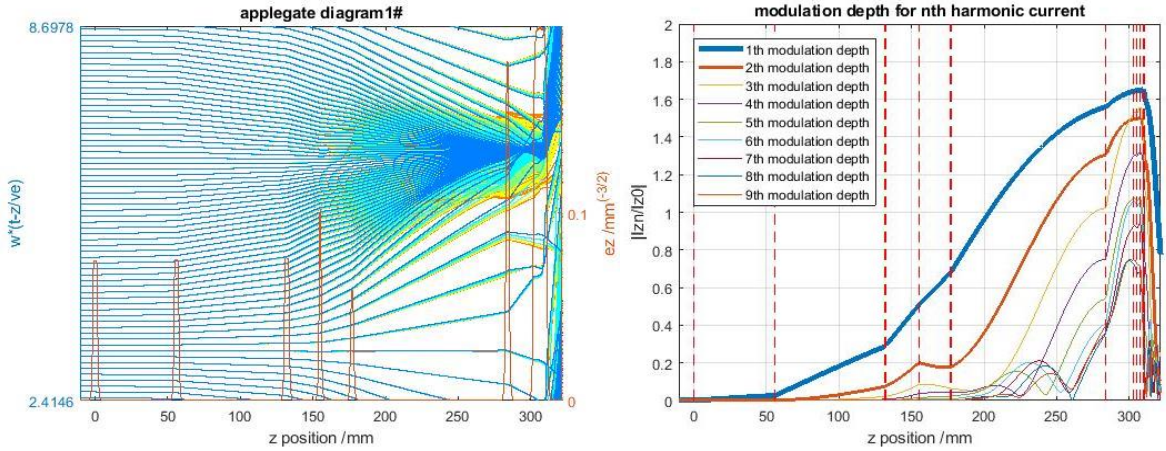

Figure 8: Applegate diagram (left) and Modulation depth for the harmonic currents (right).

Table 1: Harmonic current ratio

| Order | Ratio | Order | Ratio | Order | Ratio |
|---|---|---|---|---|---|
| 1st | 1.00 | 4th | 0.91 | 7th | 0.67 |
| 2nd | 0.94 | 5th | 0.75 | 8th | 0.56 |
| 3rd | 1.00 | 6th | 0.75 | 9th | 0.51 |
| Ideal 1-2-3 | 1.00 | Ideal 4-5-6 | $\pi/4 = 0.78$ | Ideal 7-8-9 | $\pi/5 = 0.63$ |

Table 2: Possible power outputs [MW]

| Order | [MW] | Order | P[MW] | Order | P[MW] |
|---|---|---|---|---|---|
| 1st | 29.8 | 4th | 27.1 | 7th | 20.0 |
| 2nd | 28.0 | 5th | 22.3 | 8th | 16.7 |
| 3rd | 29.8 | 6th | 22.3 | 9th | 15.2 |
| Ideal 1-2-3 | 29.8 | Ideal 4-5-6 | 23.2 | Ideal 7-8-9 | 18.8 |

| f0(MHz) | R/Q (Ω) | M | Qe | Qin | z (mm) |
|---|---|---|---|---|---|
| 12000 | 110.4116 | 0.9617 | 52 | 4.9469e+03 | 0 |
| 12061 | 110.8251 | 0.9614 | 100000 | 4.9537e+03 | 56 |
| 12310 | 114.3329 | 0.9592 | 100000 | 4.9726e+03 | 132 |
| 23888 | 26.6842 | 0.8921 | 100000 | 1.9557e+03 | 155 |
| 12283 | 33.9597 | 0.9565 | 100000 | 2.7346e+03 | 177 |
| 3.6615e+04 | 108.2822 | 0.6832 | 100000 | 3.8807e+03 | 284 |
| 3.5478e+04 | 109.4702 | 0.6928 | 100000 | 3.8329e+03 | 303 |
| 3.5478e+04 | 109.4702 | 0.6928 | 100000 | 3.8329e+03 | 305.3400 |
| 3.5478e+04 | 109.4702 | 0.6928 | 100000 | 3.8329e+03 | 307.6800 |
| 3.5478e+04 | 109.4702 | 0.6928 | 14.5000 | 3.8329e+03 | 310.0200 |

Simulation results summary

| | | | | | | |
|---|---|---|---|---|---|---|
| Pout= | 2.985e+04 | kW | Gain= | 47.76 | dB | |
| Eff.RF= | 66.23 | % | Eff.Bl= | 62.19 | % | |
| Re.RF= | 0.0003939 | | Re.El= | 3.297e-05 | | |
| |J1/J0|.i= | 1.582 | | |J1/J0|.o= | 1.67 | |
| ve/c.min= | -0.08211 | | |Gama|= | 0.9516 | |
| Successful iteration | Yes | | pha.s= | 52.1 | ° | |
| Reflected electrons | No | | Tcpu= | 128.7 | min | |

| |Vg|(kV) | phi(d.)/E kV/mm |
|---|---|
| 4.6765 | 4.7981 |
| 29.4754 | 30.2458 |
| 62.0670 | 63.1317 |
| 47.3844 | 47.8322 |
| 46.8492 | 27.5014 |
| 218.6080 | 92.2218 |
| 333.8628 | 152.4854 |
| 421.0524 | 192.3063 |
| 647.1491 | 295.5692 |
| 317.5233 | 145.0201 |

**Figure 9: KlyC set-up interface (up) and simulation output (down) using Superfish-computed Eigen-frequencies.**

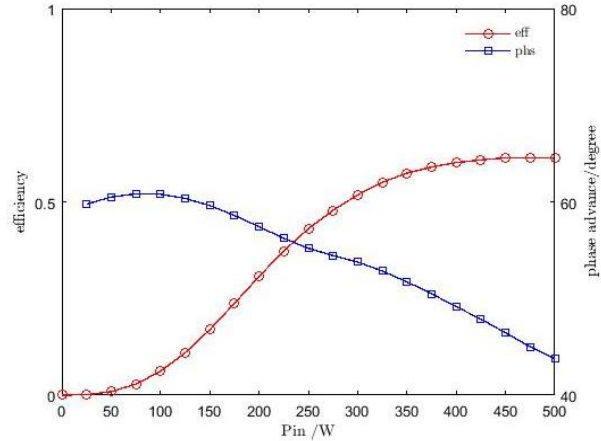

**Figure 10: Efficiency behavior in function of the input power.**

## 5. Design validation and measurements for feasibility analysis

The simulations performed by the code KlyC with the assistance of the developers has allowed to validate the design by analysis. The code KlyC has been validated by several test and its results crosschecked with other commercial and industry-reserved simulation codes including MAGIC [14] and Thales Klys-2D [15].

In order to assess the feasibility of the structure, cold measurements on sample cavities and high gradient design criteria have been analyzed. The first aspect that has been verified is the performance obtainable in terms of figures of merit of the nose-cone and pill box-cavities used in X band and Ka band.

The X band cavities used in the simulated structure have the same profile of the resonators manufactured for the accelerating structure of longitudinal phase space linearizer for the Coherent Light Source (SPARC) in Frascati. The tested oscillator is composed by 9 cells operating on the π standing wave mode [22] [23]. The structure has been manufactured by standard high precision CAM milling and assembled without brazing with a torque of 5 Nm, corresponding to a pressure of circa 80 N/mm$^2$. The structure has been measured with a network analyzer Agilent N5230A [24]. The central cavity is 13.12 mm long and it integrates a coupling aperture that is 4 mm large connected to a standard X-band waveguide by a waveguide-tapered line. At the nominal resonant frequency of the whole system (cells and coupler) f= 11.424 GHz, the measured return loss is |S11|dB = -15 dB which is in agreement with the simulation results and satisfy the design requirements.

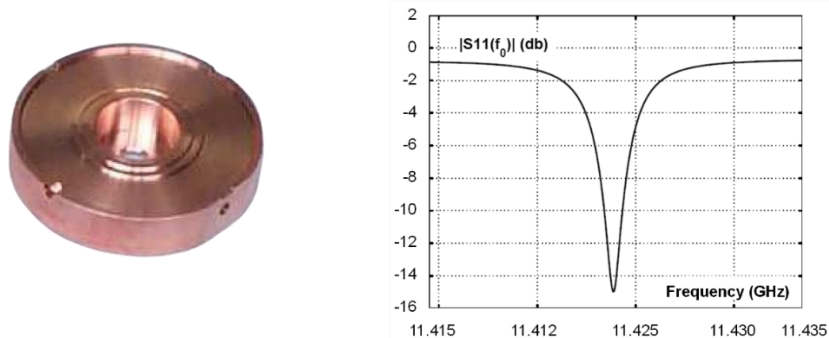

**Figure 11: Manufactured cavity section (left) and measured reflection coefficient |S11| (right).**

The Ka band cavities used in the simulated structure have the same profile of the resonators manufactured for the single–feed mode launcher (ML) dedicated to the accelerating structure of the CLIC at the CERN. The elements of the CLIC accelerator are made of OFE Cu, the roughness less than 0.4 μm and ±15 μm in shape. To achieve these tolerances, high speed milling using diamond tools and intermediate stress relieve heat treatments [25] .The mode launcher is an ensemble of a first element (the converter) that launches a wave pattern in the circular waveguide and a cavity (the transformer cell) that provides the matching with the accelerating structure [26],[27]. The structure has been manufactured, brazed and measured at the CERN [26]. At the nominal resonant frequency of the whole system (converter and transformer) f= 30 GHz, the measured return loss is |S$_{11}$|$_{dB}$ = 50 dB which is in agreement with the simulation results and design requirements.

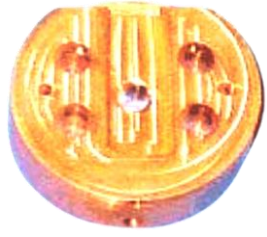 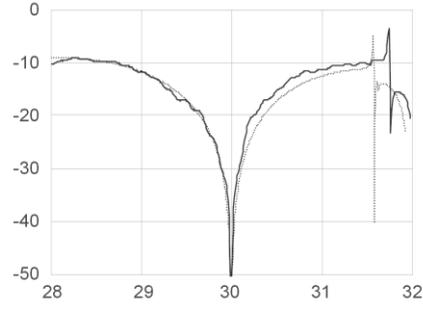

**Figure 12: Manufactured cavity section (left) and measured reflection coefficient |S11|dB (right).**

The second aspect that has been analyzed is the compliance with the main parameters for vacuum RF breakdown prediction on high gradient structures: The maximum surface electric field $|E|_{max}$ and modified Poynting vector $S_c$ as well as the surface temperature increase $\Delta T$ [18], [19] have been analyzed in case of full power driving of the case study structure. As it can be remarked in section 1.1, the maximum level of electric field that has been imposed as design constraint (300 MV/m) was already higher than the criteria on maximum electric field assumed for actual normal conducting structures (250 MV/m) [18]. This choice relies on the scope of this paper that would demonstrate the power of the proposed closed formulas and design technique to speed up the design process and get good structure addressing the limit of actual frontier. For this reason, criteria on the max surface electric field and the modified Poynting vector are overcame by 20% and 30% respectively.

The robust behavior of the proposed structure, in function of the input power (Figure 10), allow to respect all these criteria by simply reducing the input power without any need of redesign or adjustment and while obtaining high power and efficiency. Namely, by using an input power of Pin = 290 W it is possible to achieve Pout=24 MW and η = 50 % while ensuring $|E|_{max}$= 248.6 MV/m and $S_c$ = 4.56 W/µm².

**Table 3: Main parameters for breakdown prediction at the two working points.**

| $n_{cav}$ | Parameters for Pin=500W, Pout= 29.85 MW, η=62.2% | | | Parameters for $P_{in}$=290W, $P_{out}$= 24.13 MW, η=50.3% | | |
|---|---|---|---|---|---|---|
| | $|E|_{max}$ [MV/m] | ΔT [°C] | Sc [W/µm²] | $|E|_{max}$ [MV/m] | ΔT [°C] | Sc [W/µm²] |
| 1 | 4.86 | 0.00 | 0.001 | 2.44 | 0.00 | 0.000 |
| 2 | 30.57 | 0.02 | 0.021 | 15.88 | 0.01 | 0.006 |
| 3 | 63.67 | 0.06 | 0.081 | 34.63 | 0.02 | 0.024 |
| 4 | 48.07 | 1.06 | 0.212 | 30.62 | 0.43 | 0.086 |
| 5 | 27.61 | 0.12 | 0.048 | 22.39 | 0.08 | 0.032 |
| 6 | 93.48 | 4.89 | 0.788 | 62.95 | 2.22 | 0.357 |
| 7 | 154.86 | 8.86 | 1.768 | 126.60 | 5.92 | 1.181 |
| 8 | 195.30 | 14.08 | 2.812 | 152.89 | 8.63 | 1.723 |
| 9 | 300.17 | 33.27 | 6.642 | 248.59 | 22.82 | 4.555 |
| 10 | 147.28 | 8.01 | 1.599 | 131.91 | 6.42 | 1.283 |
| Criteria | < 250 | < 50 | < 5 | < 250 | < 50 | < 5 |

As reported in [28],[29],[30],[31],[32],[33], several cavities of similar dimensions operating in these bands have been manufactured and tested while showing compliance with main parameters for vacuum RF breakdown prediction on high gradient structures and manufacturing feasibility by using traditional milling techniques. The results of the feasibility analysis give full confidence in the proposed design formulas and procedure giving a solid basis for the future implementation of the presented structure.

## 6. Conclusions

This paper proposes a novel Klystron structure to implement a frequency multiplication of the drive frequency able to provide an output signal at a specific harmonic frequency. A new design principle is used proposing new analytical formulas and a specific design procedure.

A design strategy is proposed to generate a signal compression on opportune elements of the bunching circuit to supply the multiple mode excitation of the output circuit to make the beam interact with two oscillations: one of them responds to the modulations produced by a particular gain harmonic cavity the other lead the output

12arXiv:2212.12359v7 [physics.acc-ph] 23 Dec 2022

signal. The proposed principle has been described by means of a topologic structure and a detailed analytical formulation and finally and verified by a numerical modeling developed on a case study applied to the Ka-band.

The numerical model has been applied to the needs of the Compact Light XLS project, to design a high power Klystron multiplier optimized to provide a 36 GHz output while receiving a 12 GHz input. The proposed structure can be supplied by same low lever RF driver of an X-band klystron, offering the possibility to control the phase of the two power source outputs. Lower parasitic and phase noise than using a higher frequency driver are also present. With an input power Pin=500W and an external quality factor Qext=52 at the input cavity, the proposed structure shows an efficiency of η=62.2 % and a gain of 47.8 dB with less than the 5% of Ohmic losses. At the output of the device, a signal at the 3$^{rd}$ harmonic of the drive frequency is generated with a power of 29.9 MW. The max gap voltage is 647.1 kV and the max electric field is 295.6 MV/m. The minimum velocity level is -0.082·c, which respect the common design requirement on $v_{min}$ > -0.1·c, to prevent electron reflections.

The proposed design procedure and closed form formulas are applicable to other bands or power levels. As for the case study, a frequency multiplier using a drive frequency at 4 GHz can provide, by considering the reduced difficulties related to the lower frequency bands, can be used for generating a 8GHz or 12 GHz signal as well as 16 GHz or 20 GHz, by exploiting the 2$^{nd}$ or 3$^{rd}$ as well as the 4$^{th}$ or 5$^{th}$ harmonic. At the same time, a 3 GHz drive frequency ca excite a 4$^{th}$ harmonic system to generate a 12 GHz response. The proposed design can be implemented as a multi-beam klystron: by using 10 beams a power output of ~300 MW can be provided at 12 GHz, 24 GHz or 36 GHz, while maintaining the ~60% efficiency. A power of 152 MW at 108 GHz can be possible with 9$^{th}$ harmonic circuit.

In order to extend the applicability to higher frequencies of the proposed principle, or to reduce machining complexity, structures using higher order mode cavities can be engineered. Cavity prototypes have been tested through several measurements in order to show the feasibility of the proposed structure in terms of manufacturing capability and vacuum breakdown at the desired pulse duration. Without need of adjustments, the proposed case study structure can operate, at slightly different power levels (ranging from 24 MW to 30MW by varying the input power from 290W to 500W respectively), below or above the traditional thresholds of vacuum breakdown parameters for high gradient structures, giving the possibility to investigate the limit of new technologies.

# 7. Acknowledgements

The authors would thank Dr. Igor Syratchev and Dr. Jinchi Cai from CERN for the support on the software KlyC and for providing the measurements of Ka-band cavities as well as Dr. Antonio Falone for the measurements of the X-band cavities.# 8. Preprint publication notice

This work is being submitted to the IEEE for possible publication. Copyright may be transferred without notice, after which this version may no longer be accessible.# 9. References

[1] I. A. Guzilov et Al. "BAC method of increasing the efficiency in klystrons", 2014 Tenth International Vacuum Electron Sources Conference (IVESC), 30 June-4 July 2014.
[2] D. A. Constable et Al., "High Efficiency Klystron Development for Particle Accelerators", Proceedings of eeFACT2016, Daresbury, UK.
[3] A. Yu Baikov et Al., "On the Synthesis of High-Efficiency CSM Klystrons by the «Embedding» Method", 2018 International Conference on Actual Problems of Electron Devices Engineering (APEDE), 27-28 Sept. 2018.
[4] X. Chang, Y. Jiang, S. V. Shchelkunov, and J. L. Hirshfield, "Ka-Band High Power Harmonic Amplifier for Bunch Phase-Space Linearization", presented at the North American Particle Accelerator Conf. (NAPAC'19), Lansing, MI, USA, Sep. 2019, paper WEPLM51.
[5] Zhang, Haiyu et Al., "The beam-wave interaction calculation of a Ka-band Extended Interaction Klystron", IEEE International Vacuum Electronics Conference (IVEC), Monterey, USA ,19-21 April 2016.
[6] Passi, D., Leggieri, A., Di Paolo, F., Tafuto, A., Bartocci, M., "Spatial power combiner technology", Progress in Electromagnetics Research Symposium, 2015, 2015-January, pp. 932–938
[7] Leggieri, A., Orengo, G., Passi, D., Di Paolo, F., "The Squarax spatial power combiner", Progress In Electromagnetics Research C, 2013, 45, pp. 43–5513